\documentclass[12pt]{article}

\usepackage{geometry}
\usepackage{mathtools}
\usepackage{bbold}
\usepackage{hyperref}

\usepackage[dvipsnames]{xcolor}

\usepackage{algorithm}
\usepackage{caption}
\usepackage{romannum}
\usepackage{marginnote}
\usepackage{mparhack}
\usepackage{upgreek}

\begin{document}

\begin{titlepage}
   \begin{center}

       \textbf{MRI Simulation and Reconstruction Framework for Magnetic Vector Fields}\\
       \vspace{0.5cm}
       \text{Fabian Bschorr$^{1}$, Thomas Hüfken$^{1}$, Tobias Lobmeyer$^{1}$, Volker Rasche$^{1,2}$}\\
       \vspace{0.2cm}
       $^{1}$ Ulm University Medical Center, Albert-Einstein-Allee 23, 89081 Ulm, Baden-Württemberg, Germany \\
       $^{2}$ Core Facility Small Animal MRI (CF-SANI), Albert-Einstein-Allee 11, 89081 Ulm, Baden-Württemberg, Germany \\                
   \end{center}
   Abstract\\
   Purpose: Conventional MRI is relying on the assumption of the magnetic field being homogeneous in direction and amplitude. However, with the growing interest in portable, affordable point-of-care MRI systems, these assumptions do not necessarily hold anymore due to compromises necessary to achieve a reduction in e.g. footprint, weight and portability. Simulation software may help by evaluating new encoding schemes which are optimized for non-ideal hardware but also with the design of new scanner designs. The goal of this work was to develop a MATLAB-based simulation software capable of dealing with deflected magnetic fields during signal simulation and reconstruction and enabling the evaluation of arbitrary magnetic field configurations for encoding in MRI.\\
   Methods: Conventional matrix-based Bloch simulation is limited in its applicability to arbitrary magnetic fields. We therefore adapted, evaluated, and validated a modified approach, achieving substantially shorter simulation time. Furthermore, it’s used to predict image quality in 2D gradient echo experiments with deflected magnetic fields.\\
   Results: The comparison of numerical Bloch and matrix-based simulation revealed close agreement of both reconstructed images. Further, it was shown that compensation of the associated artifacts can be achieved by incorporating knowledge about the used magnetic fields into the reconstruction process.\\
   Conclusion: The presented and validated software package enables full consideration of angular inhomogeneities of magnetic vector fields used in MRI for signal simulation but also reconstruction removing the related artifacts. As such, the software might become a valuable tool for new low-field system designs and the investigation of new reconstruction algorithms.
\end{titlepage}

\section{Introduction}\label{sec1}
Magnetic Resonance Imaging (MRI) and Spectroscopy (MRS) have become flexible tools that are not only applied to anatomic assessment but are also capable of providing functional details and tissue characterization. As such, MRI/S has rendered as an inevitable tool in a broad spectrum of medical diagnosis, treatment selection, and monitoring. With the recent developments in the field of easy-to-use hyperpolarization (HP) devices \cite{nagel2023PHIP, gierse23fumarate}, MRI/S are likely to enter the field of organ/tissue specific metabolic profiling with strongly increasing demands to support personalized diagnostic and therapeutic medicine.

Current MRI/S systems are highly complex, multi-purpose imaging tools optimized for application in a broad range of fields with performance figures dictated by Neurology, Cardiology, Oncology, Orthopedics, and other medical fields. To comply with the resulting challenging demands, clinical MRI/S systems are usually high-field ($>$1\,T), provide high homogeneity over the anticipated volume of interest ($<$ 5\,ppm over a 30\,cm diameter sphere), use highly linear and powerful encoding fields ( $<$ 5\% deviation, 40\,mT/m), and strong and homogeneous radio frequency (RF) fields ($>$ 10µT). Consequently, these systems are bulky, expensive to purchase and operate, and cannot be operated under ambient conditions but need careful shielding from external RF noise.

With the expected increasing use of MRI/S in personalized medicine, simpler and lower-cost systems are needed, preferably as point-of-care (PoC) systems operated at the patient bed or even at GPs` or specialists` office. For realization, different routes have to be followed. Considering that many applications do not demand the highest spatial resolution and that in the case of HP applications, the MR signal is no longer dominated by the field strength, most likely low-field MR systems will play an important role in the future. Various low-field designs and scanners have been introduced \cite{shellock23Scanner} for dedicated applications such as brain \cite{Cooley2021PortableBrain, Liu2021Brain}, lung \cite{Tsai2008RosenHyperpol}, breast/spine \cite{Selvaganesan2023NuBoBreast} and extremity imaging \cite{Nakagomi2019Elbow, sarracanie2022Bern}. With the introduction of portable PoC systems ($\sim$65\,mT), e.g. Hyperfine Swoop (Hyperfine Research, Guilford, Connecticut, USA) or Promaxo MRI System (Promaxo Inc, Oakland, California, USA) \cite{PromaxoSatya2022}, the principal applicability of low-field ($<$100\,mT) systems for dedicated applications has been proven. 

Interestingly, most of the reported systems are scaled-down conventional systems still relying on rather conventional magnet and gradient designs aiming for a homogeneous (in amplitude and direction) main magnetic field and linear encoding fields. Even though this approach allows for almost direct translation of the well-established techniques, it will likely still limit the full flexibility in system designs rising with low-field magnets. E.g. with the now available computational power, non-linear encoding fields demanding complex reconstruction techniques may be considered. Here, inhomogeneities, e.g. caused by reducing the footprint/weight/cost of a scanner to increase portability and accessibility, might be compensated during reconstruction or even used as part of the encoding scheme, as long as the magnetic field is known \cite{Selvaganesan2023NuBoBreast, cooley2015rotMag, sarracanie2020howlow}. Even when using linear encoding fields, the increasing impact of related perpendicular components (e.g. as for concomitant gradients) needs attention, since the resulting spatially-dependent deflection of the B-field may not be negligible anymore. The impact of concomitant fields has been studied for high- \cite{BernsteinPCAngio,ZhouConCom1998,Concom25Radial} and low-field \cite{devos24concom, YABLONSKIYConcomLowField, NORRIS199033, VOLEGOV2005103, NIEMINEN2010213}. With a simplified signal model, de Vos et al. \cite{devos24concom} demonstrated a limited effect of concomitant gradients in the case of a Halbach magnet and basic MR sequences, even at low-field (50\,mT) \cite{devos24concom}, unless strong encoding fields are used. However, this impact will become increasingly prominent in more compact systems as the object becomes closer to the hardware, such as single-sided MRI systems \cite{casanova2003NMRmouse, greer2019singleside} or for non-linear encoding, e.g. using matrix gradient coils \cite{littin2018matrixcoil, juchem2020DYNAMITE} in combination with reported non-linear encoding strategies such as PatLoc \cite{Hennig2008PatLoc, Gallichan2011PatLoc}, FRONSAC \cite{Wang2016FRONSAC}, O-space imaging \cite{Stockmann2010OSpace}, or Null Space Imaging \cite{Tam2012NullSpace}. These techniques have been reported to be advantageous, e.g., for the acceleration of image acquisition \cite{Scheffler2019Spread}, the reduction of coherent aliasing artifacts \cite{Wang2015FRONSACCS}, or the tailoring of the spatial resolution achieved to a specific target region of interest \cite{Stockmann2010OSpace}.

For optimization and preliminary prediction of modified encoding approaches comprising non-homogeneous (amplitude and direction) background fields, system simulation including signal generation and reconstruction appears mandatory. 

Over recent years, multiple MRI simulators were presented with different goals \cite{KOMA, MRIlab, POSSUM1,POSSUM2, CAMINO, MRISIMUL, BlochSolver, VirtualScanner, eduMRIsim, SIMRI, coreMRI, PhoenixMR}. Some of them aim at an on-demand, cloud-based infrastructure like coreMRI \cite{coreMRI}, education and/or training purposes such as VirtualScanner \cite{VirtualScanner} or eduMRIsim \cite{eduMRIsim}, others for accelerated simulations using GPU solvers like KomaMRI \cite{KOMA} or multi-purpose simulators that also offer application-specific features such as MRIlab \cite{MRIlab}, which features, e.g., tissue modeling and exchange models or JEMRIS \cite{Stocker2010JEMRIS}. Furthermore, the latter ones also allow to simulate nonlinear encoding schemes like PatLoc \cite{Hennig2008PatLoc, Gallichan2011PatLoc}. Nevertheless, all of them mainly rely on signal simulation and reconstruction by solving the well-known Bloch equations \cite{Bloch} with the assumption of negligible perpendicular field components to the main magnetic field except consideration of concomitant fields in the phase evolution. However, with decreasing field-strength and compromises in system designs, deviations of the magnetic field from the central axis need more attention during simulation. Here, the usually made simplification of the $\vec{B}$-field still being aligned with $\hat{e}_z$ during Bloch simulation does not hold true anymore requiring adaptations of the often-used standard signal equation \cite{epstein2004inhom, NMRinhom2000, yiugitler2006permanent, arpinar2009analysis}. Since, to our knowledge, no simulator dealing with the full magnetic vector fields is currently available, a proof-of-feasibility simulation package was developed from scratch to enable simulations under full consideration of the underlying magnetic field vector.

In this manuscript, a simulation package for MATLAB (MathWorks, Natick, Massachusetts, USA) allowing the simulation of arbitrary (amplitude and direction) encoding fields, which can e.g. directly be imported from magnetic field simulation software like CST Studio Suite (Dassault Systemes, Velizy-Villacoublay, France) is presented. The package includes a Bloch equation simulation to calculate the MRI signal in a receive coil, as well as the reconstruction of the simulated data. For efficient simulation, a piece-wise constant encoding field is assumed. In this case, due to the lack of a global rotating frame, a matrix-based Bloch simulation (as done for conventional MRI simulation) is translated to arbitrary field orientations and compared to numerical Bloch equation simulations. The matrix-based approach accelerates the software while maintaining simulation fidelity. 2D gradient echo images of a Shepp-Logan and brain phantom with an underlying inhomogeneous (direction-wise) main magnetic field are presented to demonstrate the capabilities of the presented package.

\section{Theory} \label{sec:Theory}
\subsection{Signal Simulation}
This paragraph summarizes the theory of simulating the detectable MRI signal. For simplification, no discrimination between magnetic flux density and magnetic field is made and will be used interchangeably.
\subsubsection{Bloch equation}
The equation describing the motion of the magnetization vector $\vec{M}$ in a magnetic field $\vec{B}$  in MRI in the laboratory frame $(\hat{x},\hat{y},\hat{z})$ is the following Bloch equation:
\begin{equation}\label{eq:BlochEq}
	\frac{\mathrm{d}\vec{M}}{\mathrm{d}t} = \gamma \vec{M} \times \vec{B} - \frac{M_x \hat{e}_x + M_y \hat{e}_y}{T_2} - \frac{(M_z - M_0)\hat{e}_z}{T_1},
\end{equation}
in which $T_1$ and $T_2$ are relaxation constants, $\gamma$ the gyromagnetic ratio and $M_0$ the thermal equilibrium magnetization assuming that $\vec{B} = B \hat{e}_z$ such that $\vec{M}_0 = M_0 \vec{B} / B$ \cite{Bloch}. This implies that the equation depends on the coordinate system being used. If there exist magnetic field contributions perpendicular to $z$, the equation does only hold true in the local reference frame with the magnetic field being aligned along the new $z'$-axis in the coordinate system $(\hat{x}',\hat{y}',\hat{z}')$.

In case of arbitrary field orientation, a more general formulation is required to describe the motion of magnetic moments in the laboratory frame.  The relaxation terms have to be rewritten such that transverse relaxation occurs in the plane perpendicular ($\vec{M}_{\perp}$) and longitudinal relaxation in the plane parallel ($\vec{M}_{\parallel}$) to the local magnetic field. Using a vector projection of $\vec{M} = \vec{M}_{\parallel} + \vec{M}_{\perp}$ onto $\vec{B}$ allows to rewrite the relaxation terms by using the following relations:
\begin{align}
	\vec{M}_{\parallel} &= \frac{\vec{M} \cdot \vec{B}}{\vec{B} \cdot \vec{B}} \vec{B} = \frac{\vec{B} \vec{B}^T}{\vec{B} \cdot \vec{B}} \vec{M} \\
	\vec{M}_{\perp} &= \vec{M} - \vec{M}_{\parallel}
\end{align}
yielding
\begin{align}
	\frac{\mathrm{d}\vec{M}}{\mathrm{d}t} = \gamma \vec{M} \times \vec{B} - \frac{\vec{M}_{\perp}}{T_2} - \frac{(\vec{M}_{\parallel} - \vec{M}_0)}{T_1} \label{eq:BlochGenRelaxRew}\\
	= \underbrace{\left( \gamma |\vec{B}(\vec{r})| \Xi - \frac{\mathbb{1}}{T_2} + \left(\frac{1}{T_2} - \frac{1}{T_1} \right) \frac{\vec{B} \vec{B}^T}{\vec{B} \cdot \vec{B}} \right)}_{ = A} \vec{M} + \frac{\vec{M}_0}{T_1}, \label{eq:BlochGen}
\end{align}
with $\mathbb{1}$ being the identity matrix and where in the last step the cross product was rewritten in matrix form for numerical solving of the equation by using
\begin{align} \label{eq:XiDef}
	\gamma \vec{M} \times \vec{B} = \gamma \vec{M} \times |\vec{B}(\vec{r})| \hat{e}_B = \begin{bmatrix}
		0 & B_z & - B_y \\
		- B_z & 0 &  B_x \\
		B_y & -  B_x & 0
	\end{bmatrix} \gamma \vec{M} = %
	\nonumber \\ \gamma |\vec{B}(\vec{r})|
	\begin{bmatrix}
		0 & (\hat{e}_B)_z & - (\hat{e}_B)_y \\
		- (\hat{e}_B)_z & 0 &  (\hat{e}_B)_x \\
		 (\hat{e}_B)_y & -  (\hat{e}_B)_x & 0
	\end{bmatrix}
	\vec{M} = \gamma |\vec{B}(\vec{r})| \Xi \vec{M},
\end{align}
with $B_{x,y,z}$ being the components of $\vec{B}$ in the laboratory frame.


\subsubsection{Solving the Bloch equation}
An analytical solution of the Bloch equation is difficult to obtain. However, a numerical solution can be calculated by iteratively solving eq. \eqref{eq:BlochGen}. The existence of analytical solutions depend on the coefficient matrix $A$ (eq. \eqref{eq:BlochGen}). 
Here, the solution is analyzed under the assumption of piecewise constant magnetic fields. Thus, the sequence is divided into several blocks, within each of which the magnetic vector field can be regarded as constant and  
the nonhomogeneous  differential equation  eq. \eqref{eq:BlochGen} has to be solved. Since $A$ does not depend on time, a solution with the initial condition $\vec{M}(t=t_0) = \vec{M}(t_0)$ is given by
\begin{equation}
	\vec{M}(t;t_0) = \mathrm{e}^{A(t-t_0)} \vec{M}(t_0) + \int_0^{t-t_0} \mathrm{e}^{A\tau} \mathrm{d}\tau \frac{\vec{M}_0}{T_1}.
\end{equation}

For reconstruction, the relaxation constants are usually not considered, therefore, the solution of the Bloch equation may be simplified using a simple matrix exponential for propagation:
\begin{equation} \label{eq: PropExpM}
	\vec{M}(t) = \mathrm{e}^{\gamma |\vec{B}(\vec{r})| \Xi (t-t_0)} \vec{M}(t_0)
\end{equation}
Since $\Xi$ results from the cross-product (eq. \eqref{eq:XiDef}), it follows that $\Xi$ is skew-symmetric and thus normal and diagonalizable such that the matrix exponential can be rewritten by using the eigenvectors $\vec{v}_j$ and eigenvalues $\hat{\lambda}_j \in \left\{-i, 0, i\right\}$ of $\Xi$:
\begin{equation} \label{eq: MpropEV}
	\vec{M}(t) = \sum_{j = 1}^3 \mathrm{e}^{\gamma |\vec{B}(\vec{r})| t \hat{\lambda}_j} (\vec{v}_j^H  \vec{M}(t_0)) \vec{v}_j
\end{equation}
similar to \cite{devos24concom}.

\begin{figure}[htbp]
	{%
		\centering
		\includegraphics[width=0.5\linewidth]{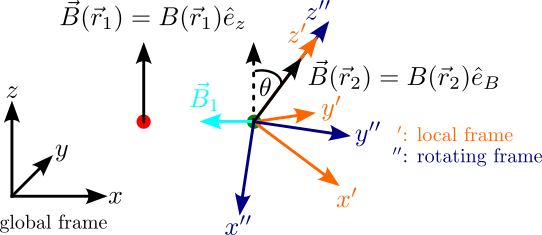}
		\captionof{figure}{%
			Coordinate systems for generalizing the MRI framework for arbitrary magnetic vector fields.}
		\label{fig:CoordArb}
	}
\end{figure}

During excitation, the angle between $\vec{B}_0$ and $\vec{B}_1$ may vary spatially. As such, a global rotating frame is not applicable, necessitating the consideration of local rotating frames. Since we are not considering strong coupling effects or very low-fields, the longitudinal components of $\vec{B}_1$ can be omitted \cite{MalcolmSpinDyn2008}. Then, the magnetization vector after a piecewise constant pulse can be calculated by transforming into a local rotating coordinate system whose $z$-axis aligns with the local $\vec{B}$ - $(\hat{x}', \hat{y}', \hat{z}')$ in fig. \ref{fig:CoordArb}. For each point in the laboratory frame, excitation is calculated by applying the respective rotation matrix after transformation into the local rotating frame $(\hat{x}'', \hat{y}'', \hat{z}'')$.

It applies for the excited magnetization vector in the global coordinate system
\begin{equation} \label{eq: ExcMat}
	\vec{M}_{\text{ex}} = \mathcal{O}_{\theta}^{-1}\mathcal{R}_{z''}(\phi)\mathcal{R}_{\hat{a}}(-\alpha)\mathcal{R}_{z''}(-\phi)\mathcal{O}_{\theta} \vec{M}_{\text{init}},
\end{equation}
in which $\mathcal{O}_{\theta}$ is the coordinate transformation into the local frame, $\mathcal{R}_{\hat{a}}(-\alpha)$ the rotation about the axis $\hat{a}$ with angle $-\alpha$ simulating the effect of the applied $\vec{B}_1$ in the local rotating frame, $\mathcal{R}_{z''}(\phi)$ the transformation into the rotating frame and $\vec{M}_{\text{init}}$ the magnetization vector just before the excitation block. 

For the transformation into the local frame, the rotation axis $\hat{n} = \hat{\mathrm{e}}_{\vec{B}} \times \hat{\mathrm{e}}_z$ and angle $\theta = \arccos\left( \frac{\vec{B} \cdot \hat{\mathrm{e}}_z}{|\vec{B}|} \right)$ have to be determined such that the transformation matrix (Rodrigues' rotation formula) can now be calculated as
\begin{equation}
	\mathcal{O}_{\theta} = \mathbb{1} + \sin\left( \theta \right) \mathcal{F} + \left(1-\cos(\theta)\right) \mathcal{F}^2
\end{equation}
using the cross product matrix $\mathcal{F}$ of the rotation axis $\hat{n}$. The transformation into the rotating frame remains identical to the conventional MRI situation, i.e. a rotation about $z'$ with the respective phase $\phi$. The excitation matrix in the rotating local reference frame can be obtained analogously as the transformation into the local frame and is in its current implementation only valid for piecewise constant pulses. First, $\vec{B}_1$ has to be transformed into the local frame and projected onto the plane perpendicular to $\vec{B}$ ($\vec{B}_{1, \perp}'$) then, the rotation axis $\hat{a} = \hat{\mathrm{e}}_{B_1'}(t=0) + (\omega - \omega_{\text{RF}})/\gamma \hat{\mathrm{e}}_{z'}$ and rotation angle $\alpha = \gamma \int_0^{t_p} |\vec{B}_{1, \perp}'(t)| \mathrm{d}t$ can be used to determine the rotation matrix $\mathcal{R}_{\hat{a}}(-\alpha)$.

\subsubsection{MRI Signal calculations}
The detectable signal can be calculated via induction and so by the principle of reciprocity \cite{Hoult1976Reciprocity}. The voltage $V_c(t)$ that is induced into a receiver coil with receive field $\vec{B}_{1,c}^{\text{RX}}(\vec{r})$ can be calculated according to
\begin{equation}
	V_c(t) = -\frac{\partial}{\partial t} \int \vec{B}_{1,c}^{\text{RX}}(\vec{r}) \cdot \Vec{M}(t) \mathrm{d}^3r.
\end{equation}
The induced voltage gets amplified, filtered and mixed in a real receive chain, such that the MRI signal $S_c(t)$ finally measured results as:
\begin{equation} \label{eq.:SigSim}
	S_c(t) = \text{LP}\left(V_c(t)\right) e^{-i \omega_d t}) = \text{LP}\left(-\int \vec{B}_{1,c}^{\text{RX}}(\vec{r}) \cdot \frac{\partial \vec{M}(t)}{\partial t}  e^{-i \omega_d t} \mathrm{d}^3r\right),
\end{equation}
in which LP refers to a low-pass filter removing fast oscillations caused by demodulation of the signal with reference frequency $\omega_d$ \cite{devos24concom}. For simulation it becomes necessary to also model the RF mixer instead of simulating the signal directly in the rotating frame, since the rotating frame transformation results spatially dependent.

\subsection{Reconstruction}\label{sec:MRISigTheory}
For reconstruction, it is convenient to simplify eq. \eqref{eq.:SigSim} further. As indicated above, it is generally difficult to find a closed form for the solution of the Bloch equation which is the reason for splitting the desired sequence into blocks of piece-wise constant magnetic fields. Propagation from block to block during the readout can be described by eq. \eqref{eq: MpropEV}. Combination with eq. \eqref{eq.:SigSim} yields an equation for the induced signal in coil $c$ at the time $t_n$:
\begin{equation}\label{eq:SignalModel}
	S_c(t_n) = -i \gamma \int |\vec{B}(\vec{r})| (\vec{v}_3^H \vec{M}(t_0)) (\vec{B}_{1,c}^{\text{RX}} \vec{v}_3) \text{e}^{\lambda_3 - i \omega_d t_n} \mathrm{d}^3r ,
\end{equation}
in which $\vec{M}(t_0)$ refers to the magnetization vector prior to the readout \cite{devos24concom}.

The magnetization $\vec{M}(t_0)$ can be obtained by solving eq. \eqref{eq:BlochGen} with all relevant contributions prior to the readout. Since we are considering a piecewise-constant magnetic vector field, $\vec{M}(t_0)$ can be obtained by applying $N$ transformations (e.g. for $N$ blocks the sequence is divided into) $\mathcal{P}_j$ onto the thermal magnetization $\vec{M}_{0} = \rho(\vec{r}) \hat{m}_0$:
\begin{equation}\label{eq:ProjM}
	\vec{M}(t_0) = \prod_j^N \mathcal{P}_j \vec{M}_{0}= \prod_j^N \mathcal{P}_j \rho(\vec{r}) \hat{m}_0.
\end{equation}
The $\mathcal{P}_j$'s are either the propagation matrices of eq. \eqref{eq: PropExpM} for a block without RF pulse or the matrices from eq. \eqref{eq: ExcMat} for a block with RF pulse.

For simulation and reconstruction of the received signal of a continuous-space phantom function, the measured phantom $\rho(\vec{r})$ needs to be discretized into voxels at positions $\vec{r}_j$.
Approximation by applying a finite series expansion yields
\begin{equation}\label{eq:DiscreteObject}
	\rho(\vec{r}) = \sum_{j} \rho_j b(\vec{r}-\vec{r}_j),
\end{equation}
where $b(\vec{r})$ describes the phantom basis function and $\rho_j$ the respective amplitude \cite{Fessler2010ModelReco}. Typically, a $\delta$-like voxel function is used, which simplifies the integral in the signal equation \eqref{eq:SignalModel} such that it can be expressed as matrix product
\begin{equation}\label{eq:MatrixSignal}
	\vec{s} = E \vec{\rho},
\end{equation}
in which $\vec{s}$ is a vector containing the signals of all encoding steps and coils, $\vec{\rho}$ is the transverse magnetization vector for all voxels and $E$ is the so-called encoding matrix. In general, for $n$ timesteps, $u$ encoding steps, $c$ receiver coils, and $N^3$ voxels, the encoding matrix $E$ can be characterized by $E \in \mathbb{C}^{n \cdot u \cdot c \times N^3}$. 
For consideration of Johnson-Nyquist and sample noise in the simulated MR signal, a noise vector $\vec{e}$ is added to eq. \eqref{eq:MatrixSignal} \cite{Fessler2010ModelReco}:
\begin{equation}\label{eq:SignalNoise}
	\vec{s} = E \vec{\rho} + \vec{e}.
\end{equation}

Linear encoding schemes reduce the image reconstruction to an inverse Fourier Transform. For nonlinear encoding schemes, motivated by eq. \eqref{eq:MatrixSignal}, the image ($\vec{\rho}$) can be reconstructed by determining the inverse of the encoding matrix $E$. The existence of a unique solution of the system of equations is not guaranteed, since the system of equations is normally under- or overdetermined. Thus, a solution of the equation can be approximated in a least-squares sense:
\begin{equation}\label{eq:LeastSquaresReco}
	E \vec{\rho} = \vec{s} \Rightarrow \hat{\vec{\rho}} = \min_{\vec{m}} || \vec{s} - E\vec{m}\, ||_2^2.
\end{equation}
If necessary, this equation can be extended by some regularization parameters in case the reconstruction problem is ill-posed \cite{Fessler20OptimizationReco}. In most cases, iterative algorithms such as the (preconditioned) conjugate gradient (pcg) algorithm \cite{hestenes1952CG, Nazareth2009CG} or an Algebraic Reconstruction Technique (ART) \cite{Gordon1970ART, chen2018kaczmarz} are applied. Where for pcg, a normal matrix is required, for ART any matrix size can be chosen.

For ART, only one row of the encoding matrix has to be calculated per iteration step in contrast to pcg for which a possibly large encoding matrix has to be stored for computation (see section \ref{sec:MRISigTheory}). Even though, ART offers the opportunity to reduce random-access memory (RAM) demands, convergence may be slow \cite{chen2018kaczmarz}.

\section{Methods}\label{sec:meth}

\subsection{Description of the simulation package} \label{sec:methDescript}

The presented MRI simulation package was developed in MATLAB (MathWorks, Natick, Massachusetts USA). Solving the Bloch equation numerically can be hardware-demanding but also time-inefficient. Thus, the matrix multiplication approach is also applied to the MRI signal simulation. The concept as described in section \ref{sec:Theory} is also applied to simulate and reconstruct MR signals / images with arbitrary $\vec{B}(\vec{r},t)$ fields not longer assuming that $\vec{B}(t,\vec{r}) = B(t, \vec{r}) \hat{e}_z$.

\begin{figure*}[htbp]
	{%
		\centering
		\includegraphics[width=0.8\linewidth]{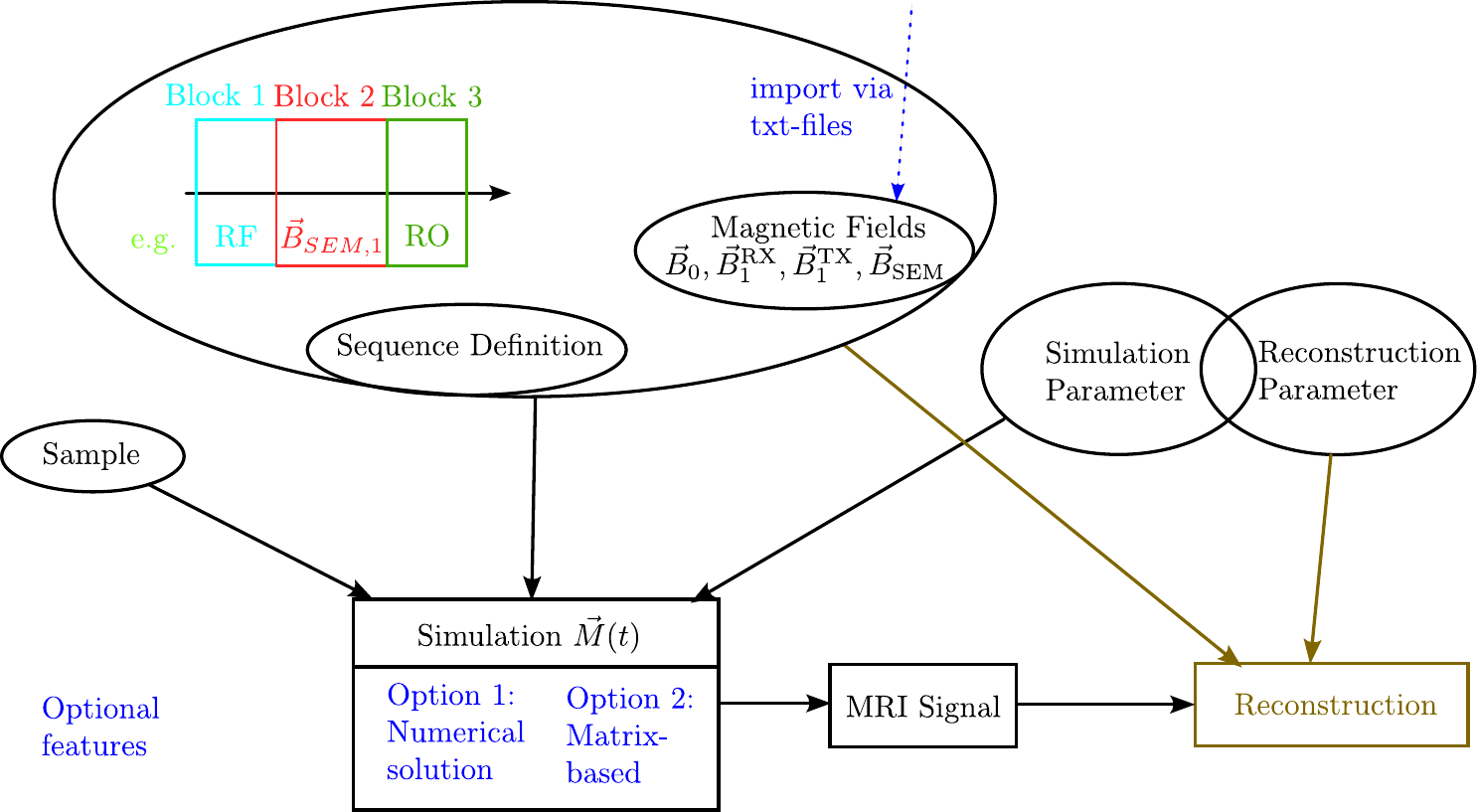}
		\captionof{figure}{%
			Graphical summary of the code structure of the presented software. Optional steps like importing the magnetic fields from a txt-file or options like using the matrix-based ansatz for solving the Bloch equation or using a numerical solution are shown in blue.}
		\label{fig:FlowCode}
	}
\end{figure*}

The code of the project is structured into four parts which are also summarized in fig. \ref{fig:FlowCode}:

\begin{itemize}
	\item[\Romannum{1}.] \textbf{Simulation parameter and sequence definition:} In addition to fundamental constants, field-of-view (FOV), signal and reconstruction matrix size, the SNR of the time-domain signal, numerical Bloch Simulation or a matrix-based simulation, the RF pulse (already implemented is a block, five-lobe sinc and gaussian pulse), and the phantom can be defined. Further, a blockwise description of the pulse sequence is defined with a block being a time interval with constant (amplitude and direction) magnetic encoding field or an excitation pulse.
	\item[\Romannum{2}.] \textbf{Definition of coordinate system and sample:} The sample can be modeled manually or an imported standard phantom is used.
	\item[\Romannum{3}.] \textbf{Simulation:}
	Based on the defined sequence, the magnetization vector $\vec{M}(t)$ is calculated by either Bloch (solving the differential equation numerically) or matrix-based (using known solutions of the differential equation to propagate the magnetization vector) simulation. The magnetic fields, i.e. static $\vec{B}_0$, transmit $\vec{B}_1^{\text{TX}}$, receive $\vec{B}_1^{\text{RX}}$, and encoding field are imported via txt-files creating a simple interface to CST. Alternatively, the desired fields can also be manually defined in the software itself. The magnetic field maps are then linearly interpolated to the requested matrix size. If using the Bloch Simulation, care has to be taken to choose the time resolution sufficiently high to capture possibly fast oscillations and reduce numerical errors. With the resulting magnetization vector, the detectable magnetization is calculated according to eq. \eqref{eq.:SigSim}.
	\item[\Romannum{4}.] \textbf{Reconstruction:} The simulated signal is reconstructed using either a preconditioned conjugate gradient algorithm or an algebraic reconstruction technique. For ART, the necessary row of the encoding matrix is only calculated per iteration, thereby reducing RAM demands by $1 / (n\cdot u\cdot c) $ using the abbreviations of sec. \ref{sec:MRISigTheory}. To account for effects introduced by non-block pulse shapes in the current version of the software, for reconstruction, additional frequency-selective terms like in \cite{pauly1989k, hennel2014effective} have to be integrated into eq. \eqref{eq:ProjM}.
\end{itemize}

The code for the software can be found at \url{https://github.com/ExCaVI-Ulm/MRISimVecField}.
To accelerate necessary simulation times, the parallelization capabilities of Matlab were used to distribute the work load onto several workers as well as a GPU-based ART implementation for reconstruction. The memory footprint of the simulation for the adapted matrix approach is mainly dominated by the definition of the magnetic field vector ($B \in \mathbb{R}^{2 \cdot t_b \cdot N \cdot 3}$ with $N$ being the total amount of voxels and $t_b$ the number of time steps to be simulated for the respective block), the simulated magnetization vector ($M \in \mathbb{R}^{t_b \cdot N \cdot 3}$) and the encoding matrix $E$.

\subsection{Validation of the Matrix Approach with Numerical Bloch Equation Simulation}
\begin{figure}[htbp]
	{%
		\centering
		\includegraphics[width=0.3\linewidth]{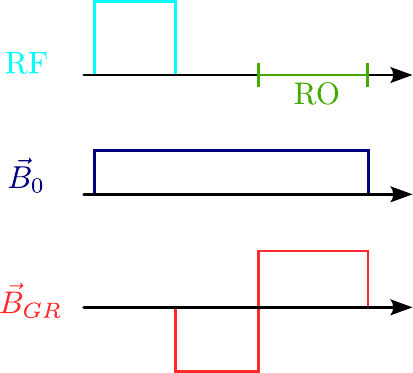}
		\captionof{figure}{%
			Sketch of the sequence used for all 2D GRE experiments.}
		\label{fig:SeqDia}
	}
\end{figure}

To validate the matrix-based approach, a comparison of a low-resolution 2D gradient echo (GRE) experiment in a $20^\circ$-tilted $\vec{B}_0$-field simulated with either the numerical Bloch equation or the matrix approach was carried out on a Shepp-Logan Phantom with matrix size $16$ and the scan parameters as shown in tab. \ref{tab:scan_param_2DGRE}. The excitation pulse was aligned with the $x$-axis as for all following experiments and $T_E = 70$\,\textmu s. To focus on the effect of angulated magnetic fields, relaxation effects were neglected and hence no specific $T_R$ is provided. 
A sketch of the used sequence is shown in fig. \ref{fig:SeqDia} whereas for the experiment described here, $\vec{B}_{GR}$ is an ideal gradient in $z$-direction and $\vec{B}_0$ an homogeneous vector field tilted by $20^\circ$ with respect to $z$. A small matrix size of $16$ was necessary to keep simulation time and especially memory-demands for the numerical Bloch simulation within reasonable limits. For good convergence in the signal simulation a high sampling frequency ($91.97$\,GHz) was used whereas for reconstruction the signal was downsampled to a lower sampling frequency ($36.79$\,MHz). The difference between both normalized reconstructed images was determined and the Frobenius norm calculated.

All simulation results shown in this publication were calculated with MATLAB R2022a on a Windows 11 workstation with a 16-core AMD CPU Threadripper PRO 5955 WX  (Advanced Micro Devices, Santa Clara, California, USA) a Nvidia GPU RTX A6000 (Nvidia Corporation, Santa Clara, California, USA) and 768 GB of RAM (Samsung Electronics, Suwon, Gyeonggi, South Korea).

\noindent\begin{minipage}{\linewidth}
	\captionof{table}{Simulation parameters for 2D GRE}    
	\begin{tabular}{ |l||l| }
		\hline
		& 2D cartesian GRE  \\
		\hline
		$B_0$ / mT & 50 \\
		FOV / mm & 200 iso \\
		Readout gradient /mT/m & 20 \\
		RF shape & block \\
		Flip Angle / $^\circ$ & 90 \\
		RF duration / $\upmu$s & 1\\
		RX Coil & 1 \\
		&with uniform sensitivity\\
		\hline
	\end{tabular}
	\label{tab:scan_param_2DGRE}
\end{minipage} 
\\ 

\subsection{Gradient Echo Imaging with Inhomogeneous $B_0$} \label{sec:GREStatic}

\begin{figure}[!htbp]
	{%
		\centering
		\includegraphics[width=0.6\linewidth]{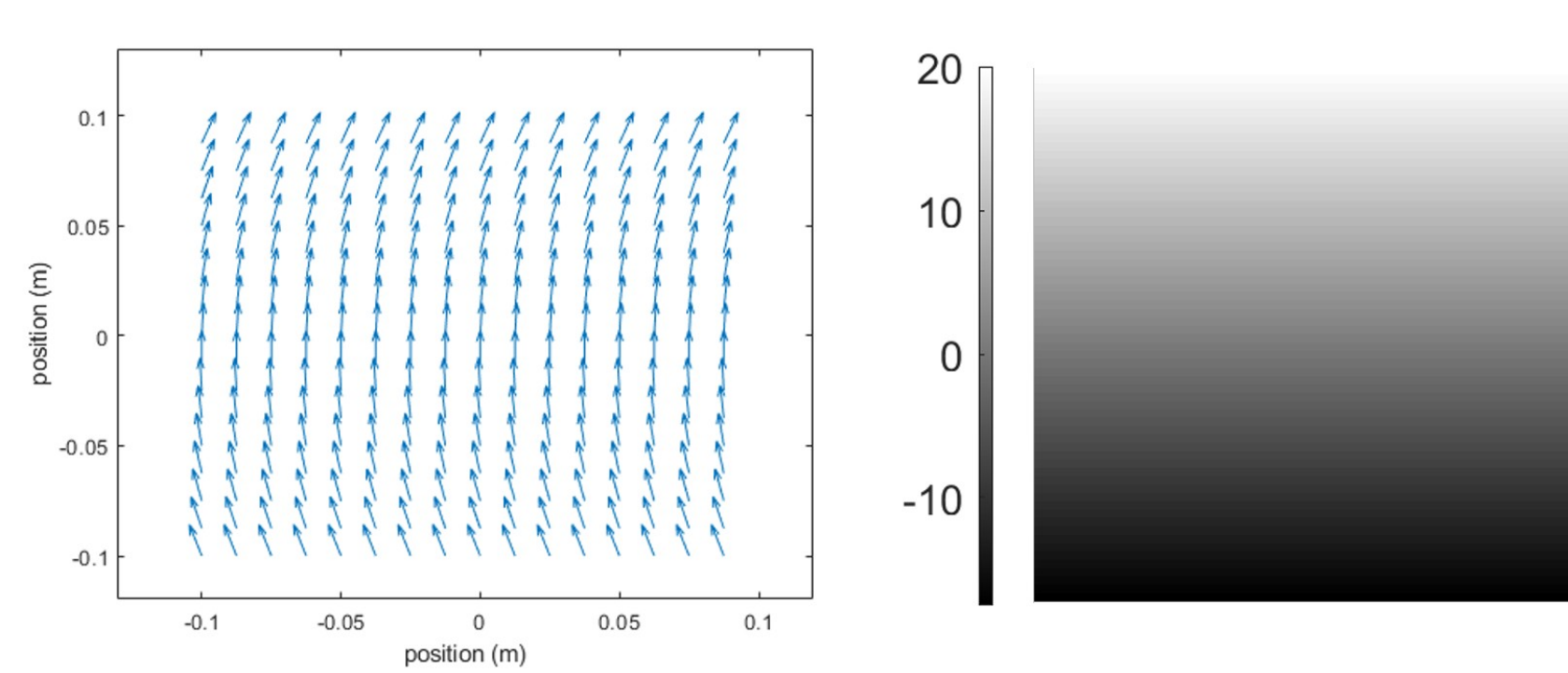}
		\captionof{figure}{%
			$\vec{B}_0$ field as used for simulation of the 2D GRE images depicted as vectors (left) and angle with respect to a vertical line (right).}
		\label{fig:GREAnglesInhom}
	}
\end{figure}

Deviations in the magnetic field directions might predominately exist in systems where the object has to be close to the hardware like suggested in the design of e.g. the NuBo scanner design \cite{Selvaganesan2023NuBoBreast}. Thus, a 2D GRE experiment with $T_E = 0.3$\,ms was simulated with an underlying $\vec{B}_0$ inhomogeneity using the sequence as shown in fig \ref{fig:SeqDia}. The angle of the magnetic field with respect to the standard MRI case, i.e. $\vec{B} = B_0 \hat{e}_z$, varied linearly in the direction up-down from $-20^\circ$ to $17.5^\circ$ as shown in figure \ref{fig:GREAnglesInhom}. The left plot shows the spatially varying direction of the magnetic vector field whereas the right plot depicts the angle of the magnetic field with respect to a vertical line as it would be for conventional MRI. The described and shown field is used as $\vec{B}_0$ as denoted in fig. \ref{fig:SeqDia}. The direction of the encoding gradients was kept along $z$. Reconstruction was carried out with the presented reconstruction technique and with standard Fourier Transform as reference. Simulation was based on a Shepp-Logan phantom with matrix size $64$. To reduce potential undersampling artifacts and limit the analysis to the effects of angular deviations of magnetic fields, phase-oversampling, and oversampling in the time domain was carefully adapted to account for the nonlinear spatially varying resonance frequencies for all following experiments. 

Furthermore, to investigate potential intravoxel dephasing effects, the same simulation was carried out with a higher number of isochromats for the simulation, namely with a matrix size of $256$, and reconstructed with a matrix size of $64$ leading to 16 isochromats per reconstructed voxel. Further simulation parameters are shown in tab. \ref{tab:scan_param_2DGRE}.


\subsection{Gradient Echo Imaging with Concomitant Gradients}

Linear encoding fields in one spatial dimension always cause concomitant fields due to Maxwell's equations. De Vos et al. \cite{devos24concom} and  Volegov et al. \cite{VOLEGOV2005103} reported the effect of concomitant fields for low-field Halbach-based and ultra low-field systems. Here, the effects of concomitant fields of an imaging system using a conventional Helmholtz-type gradient coil for a 2D gradient echo experiment were simulated with a Shepp-Logan and resolution phantom with matrix size $64$ and the reconstructed image compared to a standard Fourier Transform reconstruction. The sequence used for the simulation is depicted in fig. \ref{fig:SeqDia} with $\vec{B}_0 = B_0 \hat{e}_z$ and $\vec{B}_{GR}$ has the conventional ideal linear contribution but also the concomitant field contribution in $y$. The prephaser for the readout gradient was played out simultaneously with the phase-encoding gradient with a maximum amplitude of $50$\,mT/m to demonstrate the effect of strong encoding fields.

Further, to validate the approach and compare the results of our simulation framework to experimental data, we virtually rebuilt the tube phantom of the publication of de Vos et al. \cite{devos24concom}, i.e. 37 tubes with 15\,mm diameter and 28\,mm spacing, and used it as input for the simulation framework with $B_0 = 47$\,mT. A 15\,mT/m maximum phase-encoding and prephaser of the readout gradient was followed by a 4\,mT/m readout gradient with the concomitant fields expected from a Halbach-based system as described in \cite{devos24concom}.

\subsection{More Realistic Gradient Echo Imaging with Static Inhomogeneity}

To fully demonstrate the potential of the software framework, a 2D GRE with underlying $\vec{B}_0$ inhomogeneity as in sec. \ref{sec:GREStatic} and more realistic sequence parameters, i.e. RF pulse length $100$\,\textmu s, $T_E = 0.75$\,ms, FOV$=200$\,mm, readout gradient $10$\,mT/m, gradient ramps and using a brain phantom from JEMRIS \cite{Stocker2010JEMRIS} was simulated. Gradient ramps were modeled as an additional block with half the amplitude of the played-out gradient. For simplification, full relaxation ($T_R \leq 5 T_1$) was assumed between subsequent encoding steps.

\section{Results}

\subsection{Validation of the Matrix Approach with Numerical Bloch Equation Simulation}

\begin{figure}[htbp]
	{%
		\centering
		\includegraphics[width=1\linewidth]{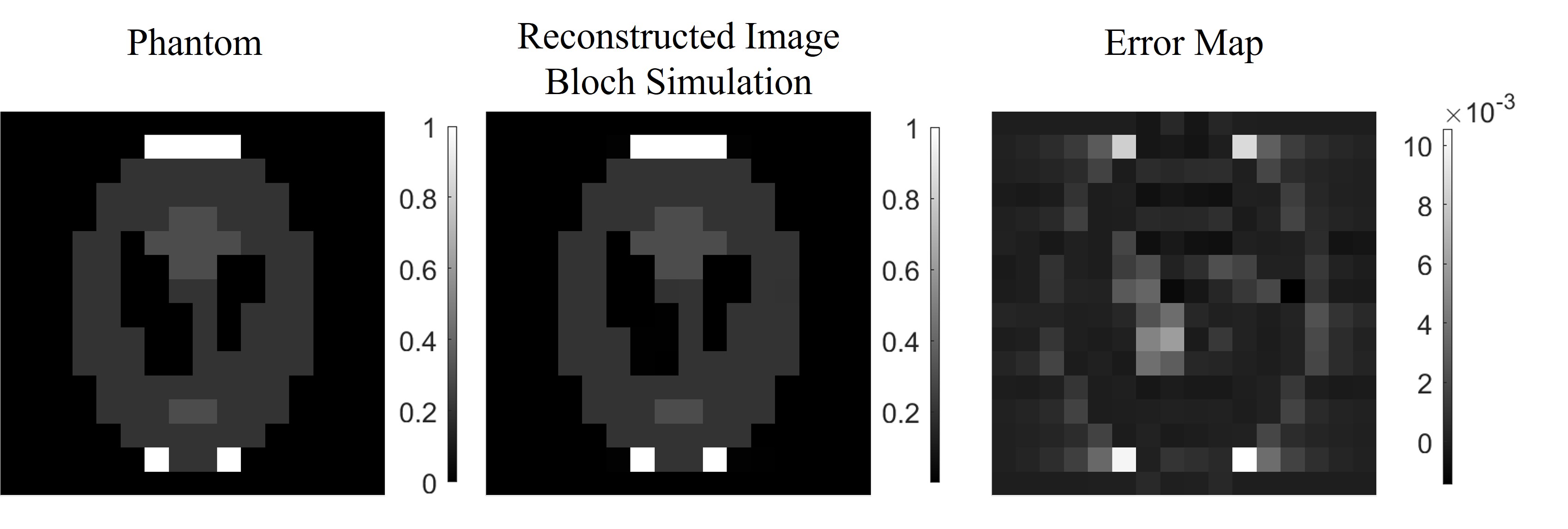}
		\captionof{figure}{%
			Validation of the matrix approach for the simulation using a Shepp-Logan Phantom (left). The reconstructed image as obtained by the numerical Bloch Simulation is shown in the middle whereas the difference of the normalized, reconstructed images from both approaches is shown on the right hand side.}
		\label{fig:CompBlochMatrix}
	}
\end{figure}

Validation of the matrix approach for the simulation was done by comparing the results with those of a numerical Bloch equation simulation. The reconstructed image simulated with the Bloch equation, as well as the used phantom and the error map, i.e. normalized difference of both reconstructed images, are shown in fig. \ref{fig:CompBlochMatrix}. The Frobenius-norm of the error map is $0.0251$, the mean $0.00059$ and the standard deviation $0.0015$ which indicates good agreement between both approaches. For this simulation, computation with the Bloch equation took $38.99$\,h whereas the matrix approach took $1.44$h, thus, reducing the necessary simulation time by about $96$\,\%. Please note, that these long simulation times result from large sampling frequencies necessary due to numerical convergence of the numerical solution. The same matrix-based simulation with reduced oversampling took about $20$\,s.

\subsection{Gradient Echo Imaging with Inhomogeneous $B_0$}
\begin{figure*}[!htbp]
	{%
		\centering
		\includegraphics[width=0.95\linewidth]{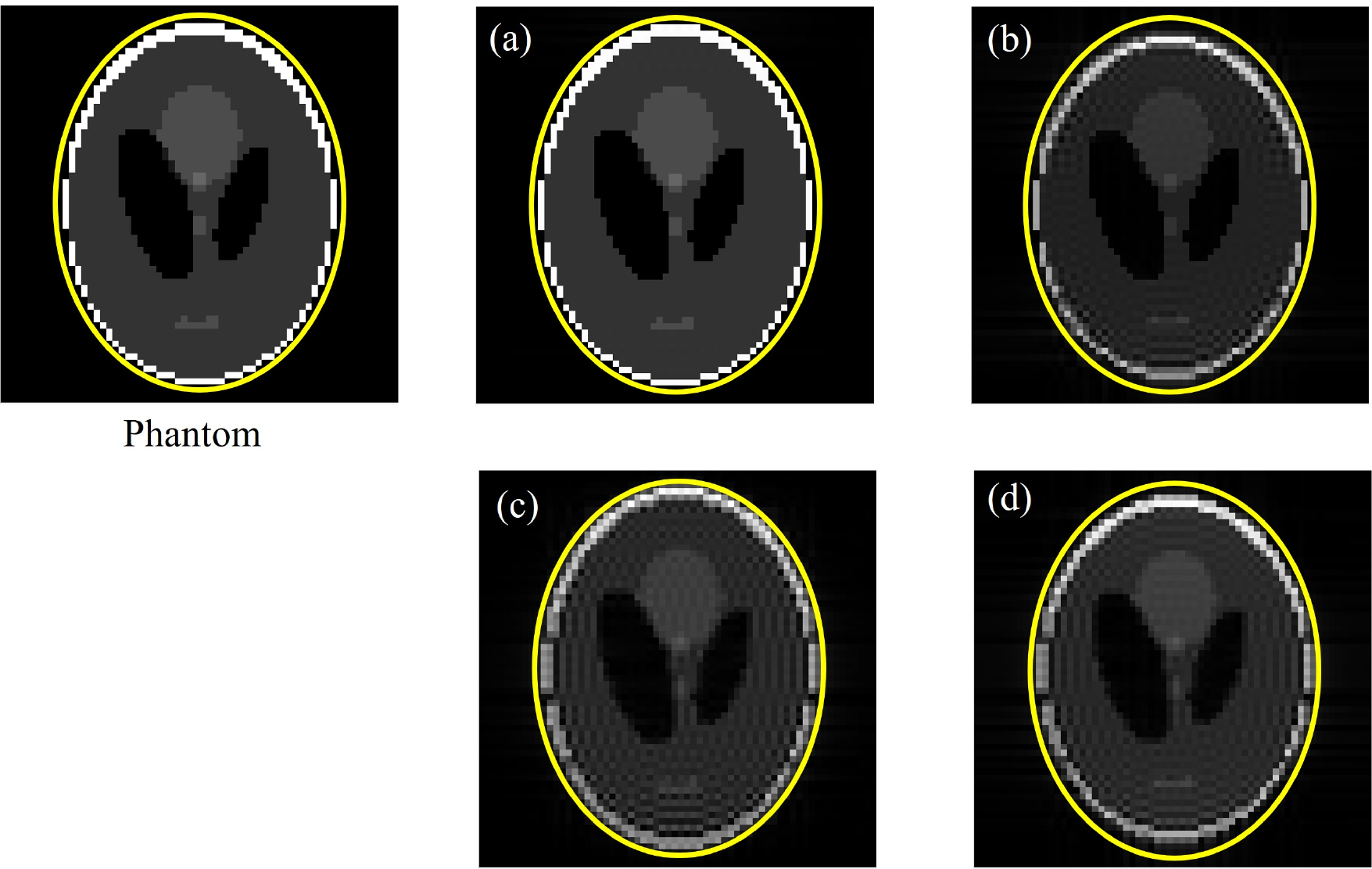}
		\captionof{figure}{%
			Simulated 2D GRE experiments with underlying $\vec{B}_0$-inhomogeneity showing the phantom, the reconstructed image (a) and a standard Fourier reconstruction (b). For (c,d), additional intravoxel dephasing was simulated for the presented reconstruction technique (c) and the Fourier-based reconstruction (d). The tilted $\vec{B}_0$ induces blurring and geometrical distortions which are visually enhanced by showing an ellipse around the phantom.}
		\label{fig:GREInhom}
	}
\end{figure*}

Fig. \ref{fig:GREInhom} shows the simulated GRE images using the aforementioned inhomogeneous $\vec{B}_0$-field. The first row of fig. \ref{fig:GREInhom}  shows the phantom as well as the reconstructed image with the presented reconstruction technique (a) and with a standard Fourier reconstruction (b). 

In the second row, simulated images, including intra-voxel dephasing, are compared between the presented reconstruction technique and standard Fourier reconstruction (c,d). For intravoxel dephasing simulation, the phantom was interpolated to a larger matrix size by next neighbor interpolation. Using the presented reconstruction corrects the apparent artifacts like blurring and geometrical distortions/warping as visible in the Fourier reconstructed images. To optically highlight the geometrical distortions which coincides with the direction in which directional deviations are simulated, an ellipse was drawn around the imaged object. As expected, adding intravoxel dephasing to the simulation adds Gibbs ringing in the reconstructed images.

\subsection{Gradient Echo Imaging with Concomitant Gradients}
\begin{figure*}[htbp]
	{%
		\centering
		\includegraphics[width=0.95\linewidth]{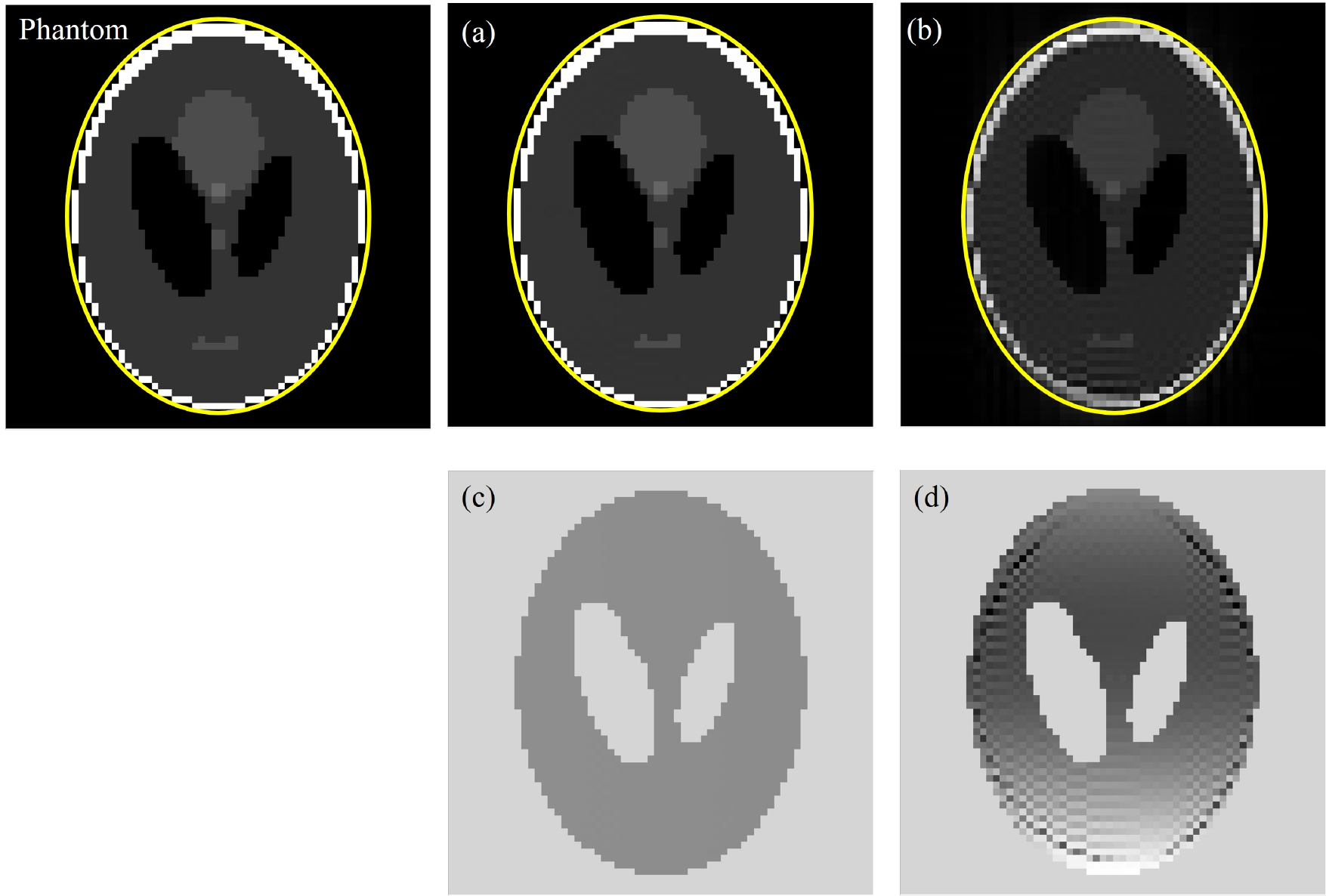}
		\captionof{figure}{%
			Simulated effects of concomitant gradient fields during a 2D GRE sequence. Compensation of the artifacts can be achieved by using the complete field information (b), also yielding an artifact-free phase (c) compared to the Fourier-based reconstructed image (b) and its corresponding phase image (d). }
		\label{fig:ConcomGRE}
	}
\end{figure*}

The simulated gradient echo images in a homogeneous, static $\vec{B}_0$-field with concomitant (dynamic) fields during phase-encoding and the readout are shown in fig. \ref{fig:ConcomGRE}. The effect of the perpendicular concomitant fields is clearly visible in form of blur and phase errors in the Fourier reconstructed image (b,d). Using the complete information about the magnetic vector field allows for compensation of these artifacts as demonstrated in (a,c). 

\begin{figure*}[htbp]
	{%
		\centering
		\includegraphics[width=0.95\linewidth]{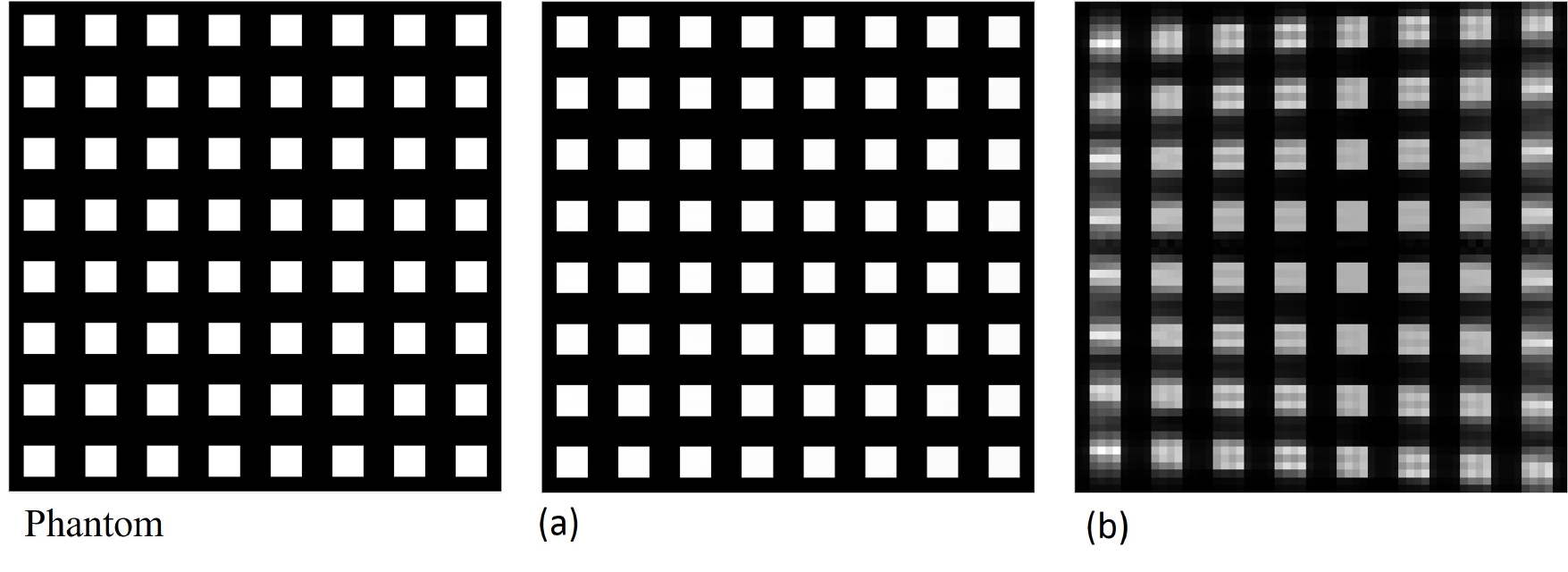}
		\captionof{figure}{%
			Simulated effects of concomitant gradient fields during a 2D GRE sequence using a resolution phantom (left image). Compensation of the artifacts can be achieved by using the complete field information (a) compared to the Fourier-based reconstructed image (b). }
		\label{fig:ConcomGREPhantom}
	}
\end{figure*}

\begin{figure}[htbp]
	{%
		\centering
		\includegraphics[width=0.35\linewidth]{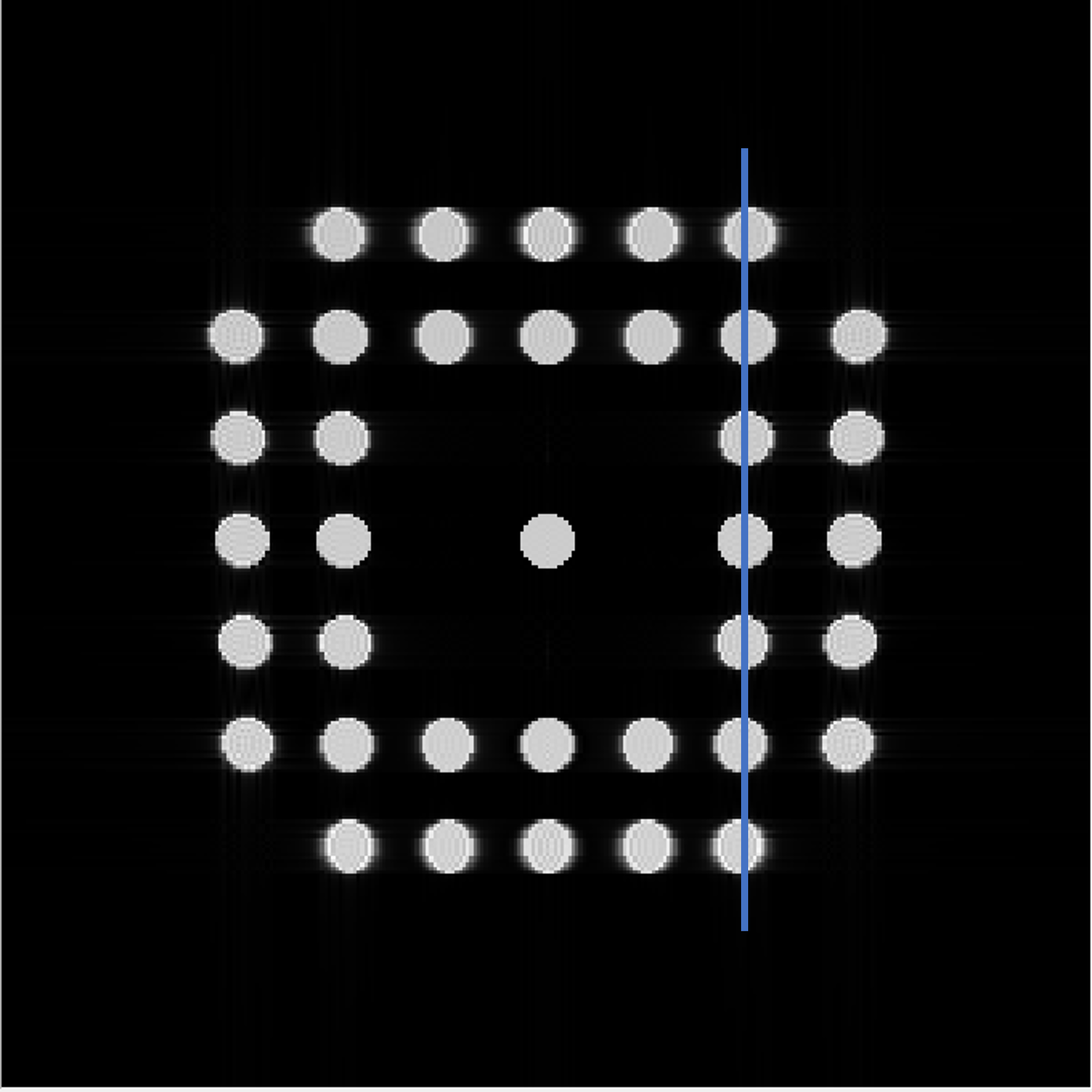}
		\captionof{figure}{%
			Simulation result of concomitant fields in a Halbach-based system with a maximum gradient strength of 15\,mT/m as reported in \cite{devos24concom}. For visualisation, a blue vertical line was added to enhance the visibility of the warping effect. FOV $300$\,mm, Matrix 300x300.}
		\label{fig:ConcomLeiden}
	}
\end{figure}

The simulated resolution phantom (fig. \ref{fig:ConcomGREPhantom}) clearly reveals the expected additional image distortion as previously reported \cite{devos24concom, VOLEGOV2005103}, which can be completely solved by the presented reconstruction fully considering the magnetic vector field. The simulation result for the comparison with experimental data obtained with a Halbach-based system is shown in fig. \ref{fig:ConcomLeiden}. The expected displacement using eq. (22) deduced in \cite{devos24concom} is 1.5\,mm, whereas the maximum displacement measured with respect to the phantom is $1$\,mm. 

\subsection{More Realistic Gradient Echo Imaging with Static Inhomogeneity}
\begin{figure*}[!htbp]
	{%
		\centering
		\includegraphics[width=0.95\linewidth]{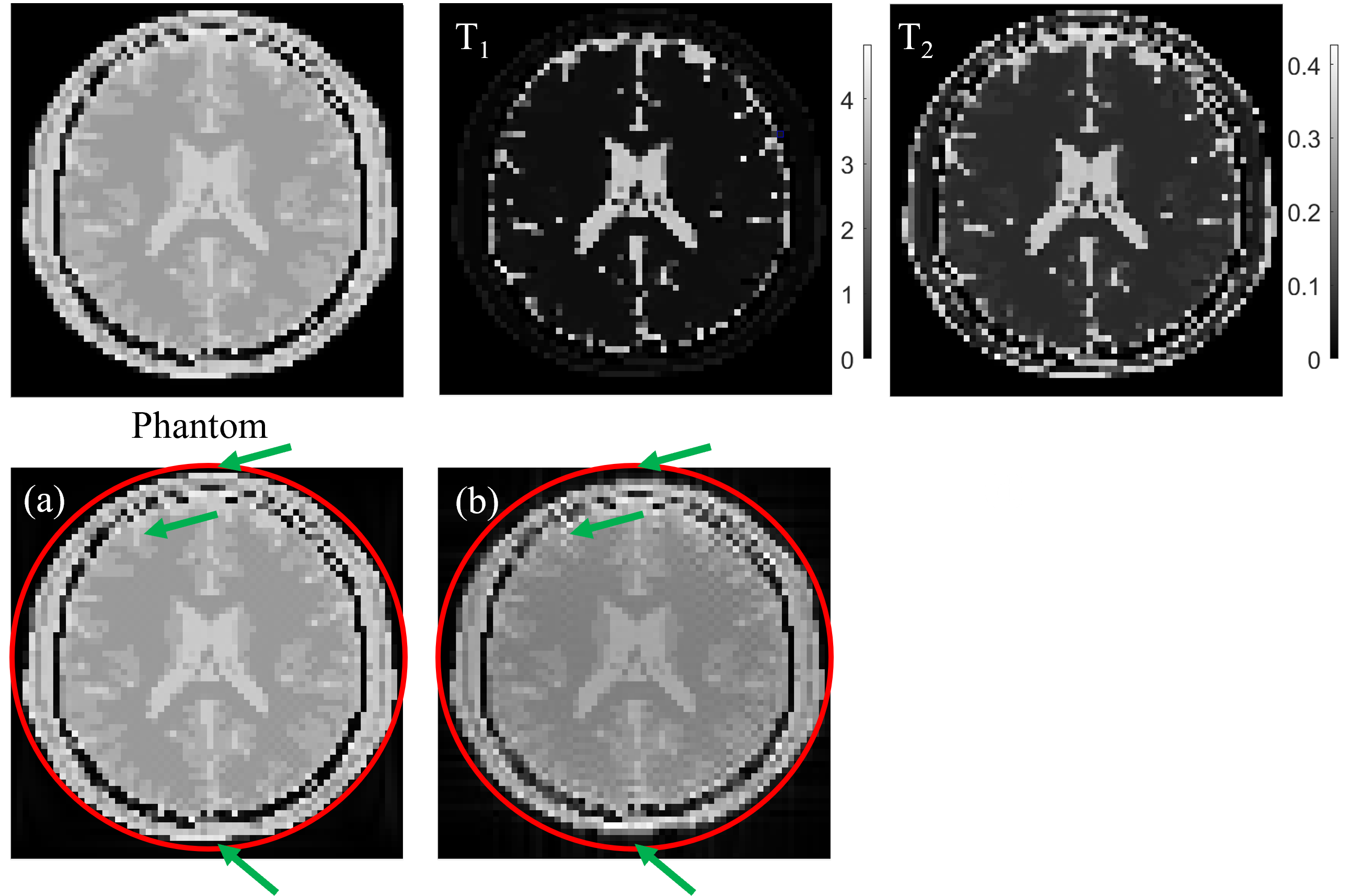}
		\captionof{figure}{%
			Simulated, more realistic 2D GRE experiments with underlying $\vec{B}_0$-inhomogeneity showing a brain phantom its $T_1$- and $T_2$-values, the reconstructed image (a) and a standard Fourier reconstruction (b). A ramp-up of the gradients was additionally simulated. The tilted $\vec{B}_0$ induces blurring and geometrical distortions which are visually enhanced by showing an ellipse around the phantom.}
		\label{fig:GRERealInhom}
	}
\end{figure*}

The results of the additional, more realistic simulation are shown in fig. \ref{fig:GRERealInhom} revealing again the potential to reduce the artifacts introduced by inhomogeneous magnetic fields as well as considering relaxation effects and gradient ramps. Small residual artifacts that appear like numerical inconsistencies during matrix inversion remain even after reconstruction considering all effects.

\section{Discussion \& Conclusions}

With the current trend towards low-field MR systems, compromises in hardware fidelity may further facilitate point-of-care systems. Considering likely deviations of the magnetic field from the usually made assumption of high-homogeneity in amplitude and direction, demands full consideration of the magnetic vector field for prediction of the potential performance of a new system design or encoding technique.

For investigating image artifacts, point-spread functions and other effects, dedicated MRI simulation software has proven beneficial, even in the case of conventional linear-encoded MRI. However, predicting image artifacts and encoding capabilities with non-linear encoding fields together with angular deviations in magnetic fields is even less intuitive, necessitating dedicated software packages to simulate the aforementioned effects. Thus, the presented simulation framework can serve as tool allowing to preliminarily evaluate the performance of a system prior to construction. The translated matrix multiplication approach as known from the homogeneous situation serves as an alternative to numerical Bloch simulation, enabling shorter simulation times without sacrificing simulation fidelity while keeping memory requirements in a reasonable range. Further, it could be shown that directional-related image artifacts can be simulated and compensated in the reconstruction for nearly arbitrary pulse sequences due to the piecewise-constant block design also offering the possibility for considering nonlinear fields for encoding.


The transferred matrix-based simulation approach was validated by direct comparison with full, numerical Bloch simulation on simulated low-resolution gradient-echo data. No apparent differences between both approaches were observed. Residuals in the order of $10^{-3}$ may be attributed to the numerical approach for solving the Bloch equations. Even though the numerical Bloch simulation implementation was not thoroughly optimized, the manifold acceleration observed with the matrix-approach appears still feasible while keeping the data fidelity close to the ground truth provided by the Bloch equation. 

Compensation of artifacts related to deflected magnetic fields either static or dynamic can be achieved by adequate considerations during reconstruction, as demonstrated in figs. \ref{fig:GREInhom}-\ref{fig:ConcomGREPhantom} for a 2D gradient echo sequence or in nonlinear encoding schemes without angular deviations in the main magnetic field, as demonstrated in \cite{bschorr2025shimenc}. Neglecting the additional contributions in the signal equation leads to blurring, amplitude-modulations, and warping/geometrical distortions as visible in the Fourier-reconstructed images. These artifacts are in good agreement with previous findings of the works on concomitant fields \cite{VOLEGOV2005103, devos24concom, NIEMINEN2010213} and intravoxel dephasing \cite{LATTA201044}. The images shown in fig. \ref{fig:ConcomGREPhantom} nicely demonstrate that warping effects dominate towards the edges of the FOV, which explains why those effects are not predominately visible in fig. \ref{fig:ConcomGRE}. Thus, the presented approach might become especially useful for single-sided, compact MRI systems designs or novel encoding strategies where the investigated anatomy lies closely to the hardware, possibly leading to substantial deflections of the magnetic fields. For those applications the presented theory and software might become a valuable tool for simulating the expected image quality prior to expensive construction of new MRI hardware designs, which is in our opinion made easy by the implemented interface to dedicated magnetic field simulation software like CST. Further, the simulation could also prove beneficial as a testbed for advanced reconstruction techniques that tackle the inversion of ill-conditioned inversion problems as faced with severe angular magnetic field deviations. In addition to the application in a low-field MRI setting, one can think of using it for high-field applications as well. Especially with recent findings of substantial artifacts caused by concomitant fields for long read-out radial-based trajectories \cite{Concom25Radial}, simulation and/or compensation of the related artifacts are essential for the design of optimized pulse sequences dealing with concomitant fields.

Furthermore, the simulation framework was used to reproduce the results of de Vos et al. \cite{devos24concom} with close agreement. However, de Vos et al. used a simplified/approximated version of the signal model to predict the effect of concomitant fields at low field strengths, which was not necessary for the presented framework, allowing the simulation to be used for a wider range of applications. Thus, our simulator extends the landscape of simulators by introducing the opportunity to generally take perpendicular field components into account which none, to the best of our knowledge, of the simulators does. Since all of the other software packages have a different focus and implementation, i.e. optimized for performance, education, specific applications, etc., of course, it would be an important next step to merge the presented approach into one of the existing simulators. This would require the implementation of a voxel-based coordinate transform for solving the Bloch equation in a frame where the magnetic field matches the local $z$-axis, which all simulators assume. A complete simulation of the RF receive path as necessitated by the lack of a global rotating frame including demodulation and sampling as well as a modification of the reconstruction according to eq. \eqref{eq:SignalModel}. The advanced mode of VirtualScanner could be a starting point for this modifications since it already features some important components of the RF receive chain like demodulation and sampling. Furthermore, another further step could be the integration of an interface to pulseq \cite{pulseq} allowing seamless access to the simulation framework.

There are a few limitations of the introduced approach/ software package one of which is the already aforementioned performance and/or hardware requirements of the host computer. For reconstruction, either a substantial amount of RAM is needed or alternative reconstruction methods like ART (which is already built-in) are used which don't require the whole encoding matrix at once. However, this often results in longer simulation times due to slow convergence properties of some reconstruction algorithms. Nevertheless, even with conjugate-gradient based reconstruction, it is possible that the simulation takes up to multiple hours, especially when using the Bloch simulation or large matrix sizes. Such large matrix sizes might result from simulating many isochromats as necessary for simulation of intravoxel dephasing effects. Here, we could only use 16 isochromats per voxel for the simulation due to hardware limitations. More adequate results might be obtained by increasing this number necessitating optimizing the performance of the software in a next step. This might be achievable by running the software on a large cluster and may be even more accelerated by splitting the matrices into smaller ones to distribute the workload on more workers or even GPUs. 
Another solution could be using alternative programming languages, which allow for a more efficient way of memory access and parallelization like other simulators aimed to optimize on, e.g. Koma \cite{KOMA} or PhoenixMR \cite{PhoenixMR}. The software already utilizes the parallelization features of MATLAB by distributing the workload to several workers, which, however, further increases the necessary RAM requirements. GPU-based acceleration is implemented for the reconstruction using ART, however, matrix sizes during signal simulation are currently too large to allow further GPU-based acceleration. Another option for solving the performance problem would be to transform the encoding matrix to frequency space in which the matrix is more sparse than in the time domain, thus, a reduction by 70\% of RAM consumption might be achieved \cite{gongRAMReco}.
Further, for reconstruction, it is necessary to exactly know the magnetic fields played out during the experiment. These fields could be monitored using magnetometers based on nitrogen-vacancy (NV) centers, which allow to monitor the magnetic field vector \cite{taylor2008high, clevenson2018robust}, however, further investigations are still to be done to determine which precision is necessary for proper inversion of the reconstruction problem. If the problem is ill-conditioned, inversion of the encoding matrix can become problematic such that artifact-free reconstruction of experimental data might not be achievable. Research towards more robust matrix inversion algorithms might become necessary to deal with challenging experimental conditions.
Despite the comparison of the result with the experimental data obtained by de Vos et al. \cite{devos24concom}, a setup is under development with which larger directional deviations in magnetic fields are expected to validate the outcomes of this work. 

In summary, we presented an open-source validated MATLAB simulation package that is capable of incorporating deflected magnetic fields into the simulation and reconstruction of MRI signals. Further, there is no limitation on using conventional encoding sequences, in principle, the software is capable of evaluating encoding capabilities of arbitrary magnetic field configurations. An interface to magnetic field simulation software is easily possible. This might pave the way for new non-linear encoding strategies that further accelerate image acquisition, for example, in the field of hyperpolarization, in which fast imaging is essential to obtain the maximum yield from limited signals. In contrast, it might also find applications in the design of portable (low-field) MRI systems by utilizing the potential for relaxed hardware constraints through the incorporation of non-ideal fields into the reconstruction process.



\section*{Acknowledgments}
This work was funded by German Federal Ministry of Research, Technology and Space grants 03ZU1110CA , 13N16446 and 03ZU2110EA.\\
The authors would like to thank Bart de Vos for the additional insights on his research about the effect of concomitant gradient fields in Halbach Low-field systems \cite{devos24concom}. Further, the authors thank the Ulm Technology Center ULMTeC for its support. \\

\bibliography{literature.bib}
\bibliographystyle{unsrt}

\end{document}